\documentclass{mnras}
\usepackage[utf8x]{inputenc}
\usepackage[english,russian]{babel}
\usepackage[T1]{fontenc}
\usepackage{array}
\usepackage{amsmath}
\usepackage{graphicx}
\graphicspath{{EPS/}}
\usepackage{float}
\usepackage{fixltx2e}
\usepackage{xcolor}
\usepackage{hyperref}

\usepackage{tikz,xcolor,hyperref}

\definecolor{lime}{HTML}{A6CE39}
\DeclareRobustCommand{\orcidicon}{%
	\begin{tikzpicture}
	\draw[lime, fill=lime] (0,0) 
	circle [radius=0.16] 
	node[white] {{\fontfamily{qag}\selectfont \tiny ID}};
	\draw[white, fill=white] (-0.0625,0.095) 
	circle [radius=0.007];
	\end{tikzpicture}
	\hspace{-2mm}
}

\foreach \x in {A, ..., Z}{%
	\expandafter\xdef\csname orcid\x\endcsname{\noexpand\href{https://orcid.org/\csname orcidauthor\x\endcsname}{\noexpand\orcidicon}}
}

\hypersetup{pdfstartview=FitH, colorlinks=true, linkcolor=blue, filecolor=blue, urlcolor=blue}

\newcommand{\sign}{\mathrm{sign}}

\newcounter{ourcount}
\title[Acceleration of protons by black holes]{Centrifugal acceleration of protons by supermassive black hole}
\author[Istomin Ya.~N., Gunya A.~A.]{Istomin Ya.~N. $^{1,2}$\thanks{E-mail: \href{mailto:istomin@lpi.ru}{istomin@lpi.ru}}, Gunya A.~A. $^{1}$\thanks{E-mail: \href{mailto:aagunya@lebedev.ru}{aagunya@lebedev.ru}}{\orcidA{}} \\
	${1}$ P.N.~Lebedev Physical Institute, Leninsky Prospect 53, Moscow 119991, Russia \thanks{Website: \href{https://www.lebedev.ru/ru/}{www.lebedev.ru}} \\
	${2}$ Moscow Institute Physics and Technology, Institutskii per. 9, Dolgoprudnyi, Moscow region, 141700, Russia \\}

\begin{document}
	\date{}
	\pagerange{\pageref{firstpage}--\pageref{lastpage}}
	\pubyear{2019}
	\maketitle
	\label{firstpage}
	
\begin{abstract}
	
	The centrifugal acceleration is due to the rotating poloidal magnetic field in the magnetosphere creates the electric field which is orthogonal to the magnetic field. Charged particles with finite cyclotron radii can move along the electric field and receive energy. Centrifugal acceleration pushes particles to the periphery, where their azimuthal velocity reaches the light speed. We have calculated particle trajectories by numerical and analytical methods.
	The maximum obtained energies depend on the parameter of the particle magnetization $ \kappa $, which is the ratio of rotation frequency of magnetic field lines in the magnetosphere $ \Omega_F $ to non-relativistic cyclotron frequency of particles $ \omega_c $, $ \kappa = \Omega_F /\omega_c << 1 $, and from the parameter $ \alpha $ which is the ratio of toroidal magnetic field $ B_T $ to the poloidal one $ B_P $, $ \alpha = B_T / B_P $. It is shown that for small toroidal fields, $ \alpha <\kappa^{1/4} $, the maximum Lorentz factor $ \gamma_m $ is only the square root of magnetization, $ \gamma_m = \kappa^{-1/2} $, while for large toroidal fields, $ \alpha >\kappa^{1/4} $, the energy increases significantly, $ \gamma_m = \kappa^{-2/3} $. However, the maximum possible acceleration, $ \gamma_m = \kappa^{-1} $, is not achieved in the magnetosphere. For a number of active galactic nuclei, such as M87, maximum values of Lorentz factor for accelerated protons are found.
	Also for special case of Sgr. A* estimations of the maximum proton energy and its energy flux are obtained. They are in agreement with experimental data obtained by HESS Cherenkov telescope.

\end{abstract}


\section{Introduction}

	At present, it is believed that the most powerful sources of gamma radiation, such as relativistic jets of blazars, Lac, quasars and radio galaxies,  are the most efficient particle accelerators (Blandford, Meier and Readhead, \hyperlink{d1}{(2019)}). Following the classic paradigm of active galactic nuclei (AGN) (Robson, \hyperlink{d2}{(1996)}), the base of these objects contains a central object in the form of a supermassive black hole surrounded by an accretion disk; its matter accrets onto the surface of the event horizon. For radio galaxies, blazars and Lac, a part of the accreted matter does not fall onto the centre, but leaves along the axis of rotation  outwards, perpendicular to the accretion disk plane, forming a collimated plasma flow (jet). The radiation of jets is non-thermal, variable and lies in a wide range of the spectrum of electromagnetic waves, from radio to X-ray.
	A number of Cherenkov astronomy observatories of (such as VERITAS, HESS, MAGIC) over the past few years have registered sources of gamma rays of extremely high energies up beyond the TeV range. The most striking sources of such energy are M87, NGC 5128, 1ES 2344+514. In addition to extragalactic sources, a significant value of radiation energy (up to $ 10^{15} eV $) was discovered in Sgr. A*, the centre of our Galaxy (Abramowski et al., \hyperlink{d3}{2016)}.
			
	The main carriers of such high energies are both electrons with energies up to TeV units (Ghisellini et al., \hyperlink{d4}{1998)} and protons with energies up to $ \simeq 10^{21} $ eV.
	To explain the origin of particles with energies in this range, the centrifugal acceleration mechanism was proposed as the main acceleration mechanism (Rieger and Mannheim, \hyperlink{d5}{2000)}; (Rieger and Aharonian, \hyperlink{d6}{2008}). Many works considered the acceleration of electrons. However, much of electrons' energy is lost are due to synchrotron radiation, so their acceleration is not as efficient as the acceleration of protons. We are interested in the possibility of reaching the limiting energy achieved during the acceleration of protons, the main component of the cosmic ray spectrum (Ginzburg, \hyperlink{d8}{1957)}.
	The acceleration of particles in the AGN can occur both in the accretion disk and in the black hole magnetosphere (Istomin and Sol, \hyperlink{d9}{2009)}. The acceleration in the disk is due to the presence of a turbulent electromagnetic field generated by the turbulent motion of the matter of the accretion disk, which leads to accretion. In the magnetosphere, acceleration of charged particles is associated with the presence of an electric field proportional to the angular velocity of rotation of the magnetospheric plasma that is brought into the rotation by the rotating black hole (Blandford and Znajek, \hyperlink{d10}{1977)}, and its accretion disk (Blandford and Payne, \hyperlink{d11}{1982)}. Also, acceleration can be induced by shock waves and in regions of reconnection of magnetic field lines. However, the first two mechanisms are always present and are stationary or quasi-stationary, while, in general, the processes leading to the appearance of shock waves and regions with a reverse magnetic field may not appear. Here we consider only the acceleration of charged particles in the magnetosphere of a black hole. The acceleration of particles in the plasma disk by a stochastic electromagnetic field was studied in detail by Istomin and Sol, \hyperlink{d9}{(2009)}. Due to interactions of energetic particles with the photon field of the disk (the disk temperature is about 10 eV), the maximum Lorentz factor of protons cannot reach large values. In this case, the acceleration in the disk can be considered as the initial process of pre-acceleration of particles injected from disk to the magnetosphere. This work is this continuation of part of the work started by Istomin and Sol, \hyperlink{d9}{(2009)}, where the acceleration of particles in a black hole magnetosphere was considered in the vicinity the accretion disk, neglecting the toroidal magnetic field.
			
	The magnetic sield structure is described in section \ref{section2}. The equations for charged particles motion are given in section \ref{section3}, along with the approximate analytical expression for maximum energies. Section \ref{section4} presense calculations for a range of real AGNs in terms of proton acceleration. The results are discussed in section \ref{section5}.

\section{Magnetic field structure}\label{section2}

\begin{figure}
	\includegraphics[width=\columnwidth]{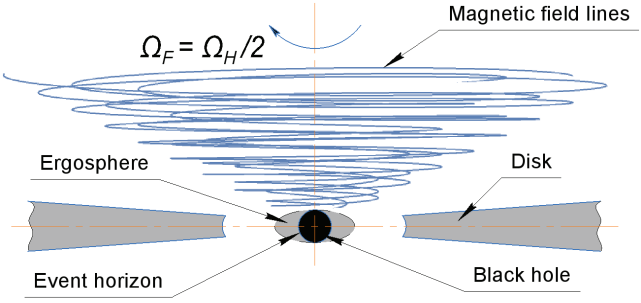}
	\caption{Scheme of picture of magnetic field lines in the magnetosphere in the vicinity of a rotating black hole}
	\label{ris:image1}
\end{figure}

	Due to rather high temperature ($T\simeq 10 \, eV$) and the relatively low plasma density of the disk
	($n\le 10^7 \, cm^{-3}$) the plasma fluid can be described
	by equations of ideal magnetic hydrodynamics (MHD). This is due to the fact that particles (protons and electrons) are magnetized, i.e.  the cyclotron frequency of protons in a magnetic field $B$,
	$\omega_{ci}\simeq 10^4(B/1\,G) \, s^{-1}$ much larger than
	frequency of proton-proton collisions $\nu\simeq 10^{-6}(n/1 \, cm^{-3})(T/1 \, eV)^{-3/2} s^{-1}$.
	Therefore,  the electric field in the disk $ {\bf E} $ is connected with the magnetic field $ {\bf B} $ by the relation
\begin{equation}\label{e}
	{\bf E}=-\frac{1}{c}{\bf u}\times{\bf B},
\end{equation}
	where ${\bf u}$ is the velocity of the disk plasma. 
	For the thin disk it means that $E_\phi = u_\rho B_z$. Here $E_\phi$ is the toroidal electric field,
	$u_\rho$ is the radial velocity of the disk matter, which does not equal to zero during accretion, $B_z$ is the component of the magnetic field orthogonal to the disk plane. The toroidal electric field
	$ E_\phi $ for stationary or quasistationary accretion is zero as the electric field is potential. Therefore, there should be no vertical magnetic field in the disk. The disk plasma should be polarized so as to push out the $ B_z $ component. Therefore at the boundary of the accretion disk the magnetic field in the magnetosphere can only have a radial, along with a spherical
	radius $ r $, component $ B_r $ and the toroidal magnetic field $ B_\phi $. By virtue of axial symmetry, i.e. non-dependence of all quantities on the azimuth angle $ \phi $, the radial field 
	$ B_r \propto r^{-2} $. It is
	the monopoly field. But due to divergence-free, the radial magnetic field must have
	different signs in different half spaces, $ z> 0 $ and $ z <0 $. Thus, $ B_r \propto s / r^2 $,
	where the value of $ s $ is the sign of $ z $, $ s = \sign(z) $. This configuration of the magnetic field is called
	the split monopoly. A toroidal field $ B_\phi $, superimposed onto the radial magnetic field, is created by radial electric currents $ j_r (j_\phi = j_\theta = 0) $, flowing along the radial magnetic field in the disk and in the magnetosphere in the immediate vicinity of the black hole. The forward and reverse currents are closed in the jet along the axis of rotation of the black hole, perpendicular to the accretion disk. A toroidal magnetic field in the general case has the form $ B_\phi\propto r^{-1} F (\theta) $, where
	$ F (\theta) $ is some function of the polar angle $ \theta $. It is determined by the dependence of the radial current $ j_r (\theta) $ on the polar angle, i.e. how the jets electric current is closed in the black hole vicinity, $ j_r \propto \partial (\sin\theta B_\phi) /
	\partial \theta / r \sin\theta $. In the simplest case, $ F (\theta) = const (\theta) $. Since the electric currents of the jet above the disk and under the disk run in different directions, the toroidal field, as well as the poloidal field, is proportional to the value of $ s $, $ B_\phi\propto s / r $. The configuration of magnetic field lines, defined by the ratio $ dr / B_r = rd \phi / B_\phi $, is expanding spirals $ r = r_1 \phi + r_0 $, where the values ​​of $ r_0 $ and $ r_1 $ are constants. Figure \ref{ris:image1} shows the magnetic field  lines of this configuration in the magnetosphere.
	
	It should be noted that the absence of the vertical magnetic field $ B_z $ in the disk refers to a stationary, or quasi-stationary, magnetic field. It means that due to the magnetization of the plasma, charged particles (protons and electrons) cannot pass through vertical magnetic field during accretion. However, if electric currents in the disk do not have time to change so that $ B_z \simeq 0 $, then a non-zero vertical magnetic field will be captured by the accreating plasma flux going to the centre and be amplified due to the magnetic flux conservation. This can lead to a decrease of the accretion rate, up to its cessation. Also can
	appear of non-stationary electromagnetic fields in the disk and particle acceleration. 
	
	However, numerical calculations of the structure of disks and their surroundings near the black hole using general relativity magneto hydro dynamics (GRMHD) show that arising inside the disk vortices strongly elongated along the disk with mainly horizontal magnetic field (Mishra et al., 2016). This means that indeed, although the field is not stationary, the electric currents in the plasma of the disk are designed so that the magnetic field created by them does not prevent accretion. For this, it is necessary that the plasma conductivity $ \sigma_c $ be small enough to suppress short vortices of the order of the disk thickness $ H $, in which the vertical component of the magnetic field is large, $ \sigma_c <v / H $. Here the velocity $ v $ is the velocity of matter in a vortex. The vortex revolution time is $ \simeq H / v $. Apparently, the numerical  magnetic viscosity (inversely to the conductivity) in the calculations satisfies this condition. Also, the real conductivity of the turbulent matter is anomalous and significantly less than the classical one. Because of this, the magnetic lines above the disk are directed tangentially to the disk (Morales Teixeira, Avara, \& McKinney, 2018) and the poloidal magnetic field in magnetosphere is close to radial one due to its strong compression by the plasma pressure of the disk near the central object (Kathirgamaraju et al., 2019; Liska et al., 2019).
	
	For split monopole magnetic field lines directly exit from the vicinity of the central black hole. They are involved into the rotation by the black hole and rotate with an angular velocity of $ \Omega_F $, which for optimal matching is half the angular velocity of rotation of a black hole
	$ \Omega_H $, $ \Omega_F = \Omega_H / 2 $ (Blandford and Znajek, \hyperlink{d10}{1977)}. 
	The rotating radial magnetic field excites in the magnetosphere a $\theta$ component of the electric field, $ E_\theta $, which, by virtue of Faraday's law, is
\begin{equation}
	E_\theta dl=-\frac{1}{c}\frac{d\Phi}{dt}=-\frac{r\sin\theta B_r}{c}\frac{d\phi}{dt}dl.
\end{equation}
	Here the element of length $ dl $ is taken along the $ \theta$ direction. We get
\begin{equation}\label{f}
	E_\theta=-\frac{r\Omega_F\sin\theta B_r}{c}.
\end{equation}
	The relation (\ref{f}) looks like a condition (\ref{e}) in ideal MHD, where instead of the
	fluid velocity $ {\bf u} $ the azimuthal rotation velocity appears, $ {\bf u} = {\bf e_\phi} r \sin \theta \Omega_F $. The electric field (\ref{f}) in the magnetosphere arises near the black hole horizon due to its rotation and is transmitted along the radial magnetic field to the magnetosphere, which is located from the centre right up to the light cylinder surface $ r = c/  \Omega_F \sin \theta $. The polar field $ E_\theta $ in the magnetosphere is the agent that transfers the black hole rotation to the rotation of plasma of the magnetosphere and, when charges are displaced in the polar direction, it transfers energy to them. We emphasize here that the polar electric field $ E_\theta $ (\ref{f}) is the result the central object rotation, and is not a consequence of MHD, which may not be satisfied in the magnetosphere. This topic is discussed in the section 'Discussion'.
	
	Thus, in the magnetosphere there is an electromagnetic field with components
	$$
	B_r=s B_0\left(\frac{r}{r_L}\right)^{-2},
	$$
\begin{equation}\label{fields}
	B_\phi=\alpha s B_0\sin\theta\left(\frac{r}{r_L}\right)^{-1}, \,
\end{equation}
	$$
	E_\theta=-s B_0\sin\theta\left(\frac{r}{r_L}\right)^{-1}.
	$$	
	Here, $ r_L = c / \Omega_F $ is the natural size of the magnetosphere in the transverse direction.
	This is the radius of the light cylinder, at which the electric field is compared with the magnetic one. For a spherical distance $ r $, the light surface is at the distance $ r = r_L / \sin \theta $.
	The value of $ B_0 $ is the value of the radial magnetic field at the distance $ r = r_L $. At the light cylinder
	the value of the radial field falls from the value of $ B_0 $ near the accretion disk $ (\theta \simeq \pi / 2) $ to
	$ B_0 \sin^2 \theta $. The coefficient $ \alpha $ is the ratio of the toroidal magnetic field to the poloidal one at the distance $ r = r_L $ from the centre. Since the split monopole field and the toroidal field differently depend on the radial distance $ r $, then at 'small' distances $ r <\alpha^{-1} r_L $ the poloidal magnetic field prevails, while at 'large' distances $ r> \alpha^{-1} r_L $ the field becomes predominantly toroidal. So for $ \alpha> 1 $ the magnetic field near the light surface, $ r = r_L / \sin \theta $, where, as we will see, the main acceleration of charged particles occurs, the field is mainly toroidal. For small values ​​of $ \alpha $, $ \alpha <1 $, the field is only at large distances from the centre, $ r> (\alpha \sin \theta)^{-1} r_L $, is close to the toroidal one, while near the field centre it is close to the poloidal one.
		
\section{Particle acceleration}\label{section3}
		
	Motion of particles of mass $ m $ and charge $ q $ in the electromagnetic field of the black hole magnetosphere is described by equations
	$$\frac{d{\bf p}}{dt}=q\left({\bf E}+\frac{1}{c}\left[{\bf v,B}\right]\right), \,$$
\begin{equation}\label{m}
	\frac{d{\bf r}}{dt}={\bf v}=\frac{{\bf p}}{m\gamma}, \, 
\end{equation}
	$$\gamma^2=1+\frac{p^2}{m^2 c^2}.$$
	Here $ {\bf r} $ and $ {\bf p} $ are the coordinate and the momentum of a particle, $ \gamma $ is its Lorentz factor.
	It is convenient for us to introduce dimensionless time, coordinates, velocity and momentum,
\begin{equation}\label{dimensionless}
	t'=\frac{\omega_c t}{\gamma_i}, \, 
	{\bf r'}=\frac{{\bf r}}{r_L}, \, {\bf v'}=\frac{{\bf v}}{c}, \, 
	{\bf p'}=\frac{{\bf p}}{mc\gamma_i}.
\end{equation}
	The initial value of the Lorentz factor is $ \gamma_i $, the nonrelativistic cyclotron frequency of a particle rotation in the $ B_0 $ field is $ \omega_c = qB_0 / mc $. Let us also introduce the value of the Lorentz factor relative to the initial energy, $ \gamma '= \gamma / \gamma_i $. In these variables, the equations of motion (\ref {m}) in spherical coordinates $ r, \theta, \phi $ (primes are omitted) have the form
\begin{eqnarray}\label{m1}
	&&\frac{dp_r}{dt}=\frac{\kappa}{r\gamma}\left(p_\theta^2+p_\phi^2\right)+\frac{s\alpha}{r\gamma}p_\theta, \nonumber \\ 
	&&\frac{dp_\theta}{dt}=-\frac{\kappa}{r\gamma}\left(p_r p_\theta-p_\phi^2 \cot\theta\right)-\frac{s}{r}\sin\theta+\frac{s}{r^2\gamma}p_\phi -\frac{s\alpha}{r\gamma}p_r, \nonumber \\  
	&&\frac{dp_\phi}{dt}=-\frac{\kappa}{r\gamma}\left(p_r+p_\theta \cot\theta\right)p_\phi- \frac{s}{r^2\gamma}p_\theta, \\  
	&&\frac{dr}{dt}=\frac{\kappa}{\gamma} p_r, \nonumber \\  
	&&\frac{d\theta}{dt}=\frac{\kappa}{r\gamma}p_\theta.  \nonumber
\end{eqnarray}
	The equations (\ref{m1}) contain two dimensionless constants, $ \alpha $ and $ \kappa $. The value of $ \alpha $, introduced in the previous section, is equal to the ratio of the toroidal magnetic field to the poloidal one at the distance $ r = r_L $, $ \alpha = B_ \phi / B_r |_{r = r_L} $. The value of $ \kappa $ is equal to
\begin{equation}\label{kap}
	\kappa=\frac{c\gamma_i}{r_L\omega_c}=\frac{r_c}{r_L}=\frac{\Omega_F}{\omega_c/\gamma_i}
\end{equation}
	and is actually the parameter of a charged particle magnetization, that is the ratio the cyclotron radius of a particle, $ r_c = c \gamma_i / \omega_c $,  to the radius of the light cylinder, or the ratio of the rotation frequency of magnetic field lines to the cyclotron frequency. It is clear that in the magnetosphere only magnetized particles can be accelerated, $ \kappa <1 $, $ \gamma_i <\gamma_0 = r_L \omega_c / c $, which conforms the Hillas criterion (Hillas, \hyperlink{d12}{1984)}, which is geometric in nature: it is impossible to accelerate a particle in a magnetic field whose scale is smaller than the cyclotron radius. 
	
\begin{figure}
	\centering{\includegraphics[width=1\linewidth]{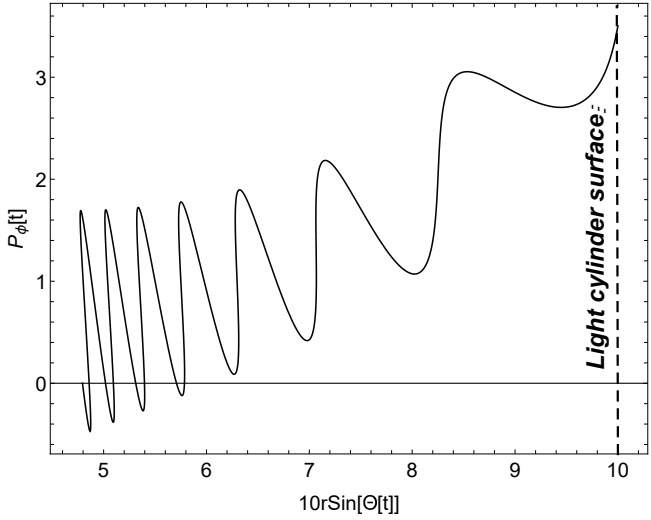}}
	\caption{The azimuthal momentum of the particle $ p_\phi $ versus the radial distance $r \sin\theta$ for $\kappa = 10^{−2}$ and $\alpha = 10^{−2}$. The light cylinder surface is located at 10 on the abscissa axis.}
	\label{ris:image2}
\end{figure}

\begin{figure}
	\centering{\includegraphics[width=1\linewidth]{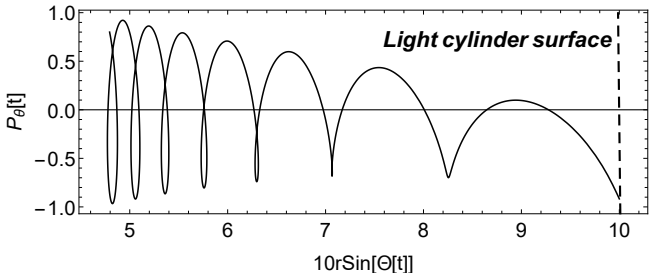}}
	\caption{The polar momentum of the particle $ p_\theta $ versus the radial distance $r \sin\theta$ for the same parameters as in Fig. 2.}
	\label{ris:image3}
\end{figure}

	The acceleration of 'cold' particles, $ \gamma_i \simeq 1 $ is considered. In this case, $ \kappa << 1 $. The question is to find the energy maximum $ \gamma_m $ when proton reaches the light surface $ r = r_L / \sin \theta $. First of all need to solve a system of equations (\ref {m1}).
	The numerical calculation of the proton trajectory starting from the magnetosphere internal region, $ r < r_L $, is show in Figures (\ref {ris:image2}) and (\ref {ris:image3}).
		
\begin{figure}
	\centering{\includegraphics[width=1\linewidth]{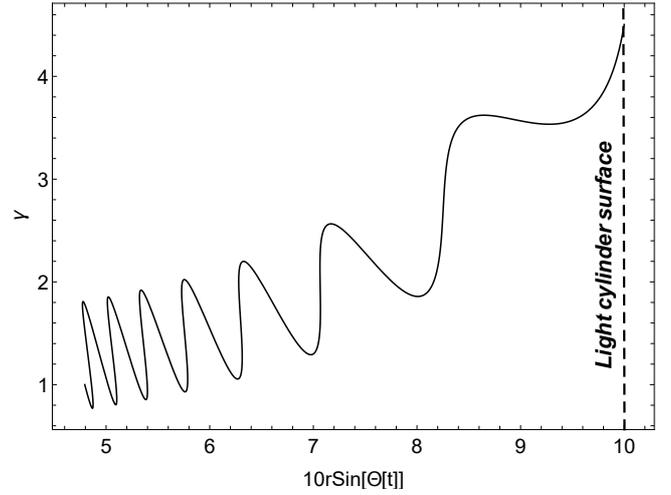}}
	\caption{The Lorentz factor of the particle $ \gamma $ versus the radial distance
	$r \sin\theta$ for same parameters as in Fig. 2}
	\label{ris:image4}
\end{figure}	
		
	First, it can be seen that the behavior of the toroidal momentum $ p_ \phi $ near the light surface, $ r \simeq 1 / \sin \theta $, practically does not differ from the behavior of the Lorentz factor $\gamma$ (energy). This is due to the fact that the system (\ref{m1}) has two integrals of motion: energy $ {\cal E} = const $ and angular momentum $ {\cal L} = const $,
\begin{equation}\label{const} 
	{\cal E}=\gamma - \frac{s}{\kappa}\cos\theta, \, {\cal L}=rp_\phi \sin\theta-\gamma.
\end{equation}
	As well as energy consisting of particle energy and the work of an electric field, the angular momentum
	is the sum of the mechanical momentum and the momentum of the electromagnetic field. Since the initial radius $ r_0 $, from which the particle starts, is less than unity, $ r_0 << 1 $, then, as it follows from the conservation of the angular momentum (\ref {const}), $ p_{\phi_{max}} = \gamma_m-1 $. The value of $ \gamma_m $ is equal to the maximum Lorentz factor of a particle achieved at the light surface.
	In addition, $ \gamma^2 = p_r^2 + p_ \theta^2 + p_\phi^2 + 1 / \gamma_i^2 $, and $ (p_r^2 + p_\theta^2) |_{max} = 2 \gamma_m \, (\gamma_m >> 1) $. Thus, particle acceleration occurs in the azimuthal direction, $ p_\phi \simeq \gamma $, while the poloidal components of a particle are much smaller, $ p_r \simeq p_\theta\simeq\gamma^{1/2} $. This is the centrifugal acceleration.
			
\begin{figure}
	\centering{\includegraphics[width=1\linewidth]{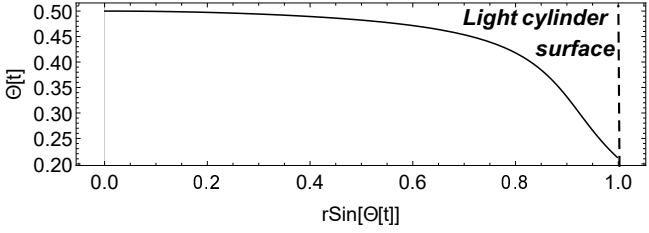}}
	\caption{The polar angle versus the radial distance $r \sin(\theta)$ for $ \kappa = 10^{-4} $}
	\label{ris:image5}
\end{figure}

\begin{figure}
	\centering{\includegraphics[width=1\linewidth]{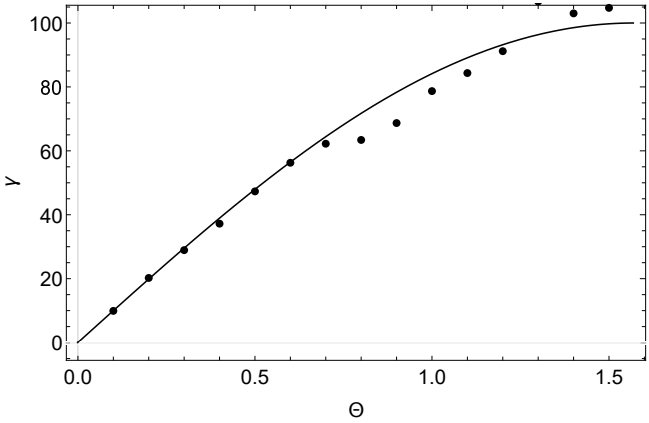}}
	\caption{The Lorentz factor of a particle at the light surface $ \gamma_m $ versus the polar angle $ \theta $ for $ \kappa = 10^{-4} $. Points are the result of numerical calculations, the curve is the dependance (\ref{alpha0}).}
	\label{ris:image6}
\end{figure}

\begin{figure}\label{ris:image7}
	\centering{\includegraphics[width=1\linewidth]{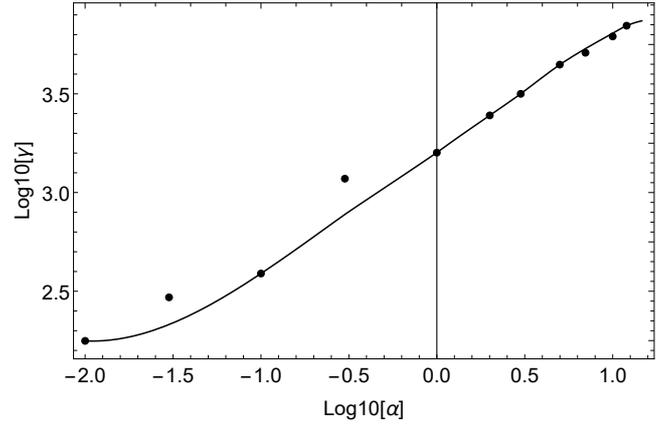}}
	\caption{The maximum Lorentz factor of particle $ \gamma_m $ versus the value of $ \alpha $ for $ \kappa = 10^{-4} $. The slope of the curve for $\alpha> 10^{-1}$ just corresponds to the dependence $\gamma_m\propto \alpha^{2/3}$ (\ref{alpha}).}
\end{figure}

	Secondly, the angle $ \theta $ varies strongly only in the vicinity of the light surface, practically without changing on the main part of the trajectory (Fig. (\ref {ris:image5})). As we will see, always $ \gamma << \kappa^{-1}, \, \kappa << 1 $, and it follows from the energy conservation law (\ref {const}) that $ \cos\theta - \cos\theta_0 << 1 $. Since the main acceleration occurs near the light surface (Fig. (\ref {ris:image4})), then $ \Delta \theta = \theta - \theta_0 = - \sigma \kappa
	\gamma_m / \sin \theta $. On the other hand, $ d \theta / dr = p_\theta \sin\theta / p_r = \Delta \theta / \Delta r $,
	where $ \Delta r $ is the size of the region near the light surface $ r = 1 / \sin \theta $, where the main acceleration of particles occurs
\begin{equation}\label{delta}
	\Delta r=-\frac{s\kappa\gamma_m}{\sin^2\theta}\frac{p_r}{p_\theta}\mid_{r=1/\sin\theta}.
\end{equation}
	Small angles $ \theta \simeq 0 $ are not considered because the possible region of acceleration in this case is far from the centre, where the fields are small and the acceleration is not efficient.
	Passing in the system (\ref {m1}) from time derivatives to derivatives over coordinate $r, \, d/dt=d/dr(p_r/\gamma)$ , and replacing derivatives by fractions $d{\bf p}/dr\simeq {\bf p}|_{r=1/\sin\theta}/\Delta r$, we obtain an algebraic system of equations that determines the values of ($\gamma_m, {\bf p})|_{r=1/\sin\theta}$ at the light surface $r=1/\sin\theta$ 
\begin{eqnarray}\label{algebra}
	&&p_r^2+\frac{\alpha\gamma_m}{\sin\theta}p_r-\frac{\gamma_m(\sin^2\theta+\kappa\gamma_m^2|\cos\theta|)}{\sin^2\theta}=0; \nonumber \\
	&&p_\theta =-p_r\frac{s\kappa\gamma_m^2\sin\theta}{\sin^2\theta+\kappa\gamma_m^2|\cos\theta|}; \\
	&&p_\theta^2+p_r^2=2\gamma_m.  \nonumber 
\end{eqnarray}
	First consider the case of small values of $ \alpha $. Put (\ref{algebra}) $ \alpha = 0 $  in the system of equations we find
\begin{equation}\label{alpha0}
	\gamma_m=\kappa^{-1/2}\sin\theta.
\end{equation}
	This result coincides with the expression obtained in (Istomin and Sol, \hyperlink{d9}{2009)} for the case of $ \theta = \pi / 2 $, where the acceleration of particles in the magnetosphere was considered only near the accretion disk. The expression $ \gamma_m = \kappa^{-1/2} (\sin \theta = 1) $ was obtained from the analysis of numerical calculations of particle trajectories, while here we use the analytical approximation.
	We show in Figure (\ref {ris:image6}) the comparison of numerical calculations (points) with the analytical expression
	(\ref{alpha0}). There are in a good agreement.
	To obtain a general expression for $ \gamma_m $ for the arbitrary value of $ \alpha $, we need to solve the quadratic equation, the first equation of the system (\ref{algebra}), with respect to the value of $ p_r $. However, the resulting expression is inconvenient for analysis, so we will find an approximate expression for the value of $ p_r $ in the case of large values of $ \alpha, \, \alpha^2 \gamma_m >> 1, $
\begin{equation}\label{pr}	
	 p_r\simeq \frac{\sin^2\theta+\kappa\gamma_m^2|\cos\theta|}{\alpha\sin\theta}. 
\end{equation}
	The result is
\begin{equation}\label{alpha}
	\gamma_m\simeq 2^{1/3}\left(\alpha\kappa^{-1}\right)^{2/3}\sin^{2/3}\theta.
\end{equation}
	By comparing  expressions (\ref{alpha0}) and (\ref{alpha}) obtain the value of $ \alpha = \alpha_1 $,
	at which the dependence of the maximum Lorentz factor $ \gamma_m $ on the magnetization parameter $ \kappa $ at small toroidal fields, $ \alpha <\alpha_1 $, transits to the dependence at large toroidal fields, $ \alpha> \alpha_1 $,
\begin{equation}\label{alpha1}
	\alpha_1 =2^{-1/2}\kappa^{1/4}\sin^{1/2}\theta\simeq\kappa^{1/4}.
\end{equation}
	Figure (\ref{ris:image7}) depicts the dependence of $ \gamma_m $ on the toroidal magnetic field magnitude.	
	The dots correspond the values calculated numerically using the equations of motion of particles in the magnetosphere (\ref {m1}). Clearly visible is the transition from independent $ \alpha $ at small values of
	$ \alpha $ relation to the relation of $ \gamma_m \propto \alpha ^ {2/3} $ in the region
	$ \alpha> \alpha_1 $.
	The result of our studies of particle acceleration shows that the acceleration depends not only on the magnitude of the magnetic field - the magnetization parameter $ \kappa $ (\ref{kap}), but also on the topology of the magnetic field, on the ratio between the poloidal and toroidal magnetic fields. 
	
	The acceleration of particles in the magnetosphere of the rotating black hole is carried out, by the electric field $ E_ \theta $ (\ref{fields}). This field arises when magnetic field lines rotate with the angular velocity of $ \Omega_F $, which is proportional to the angular velocity  of a black hole rotation. The particle, making cyclotron rotation, and drifting in crossed fields in the direction of rotation due to polarization drift (increasing its toroidal momentum) shifts from the original magnetic surface and acquires additional energy in the electric field. The strongest deviation occurs near the light surface at the distance $ \Delta r $ (\ref{delta}) from it. Here, accelerating, the particle increases its cyclotron radius and accelerates even more until it reaches the light surface. In the toroidal magnetic field near the light surface, the particle, in contrast to movement in the poloidal field,  also experiences centrifugal and gradient drifts directed along the electric field, which deviate it from the given magnetic surface and produces more efficient acceleration.
	Here the values ​​of $ \gamma $ and $ \gamma_m $ are values ​​of the Lorentz factor and the maximum Lorentz factor relative to the value of $ \gamma_i $, which is the initial Lorentz factor.
	Actually expressions
	$$\gamma_m\simeq\kappa^{-1/2}=\gamma_0^{1/2}, \,
	\alpha<\gamma_0^{-1/4}; \, 
	$$
\begin{equation}\label{gammam}	
	\gamma_m\simeq\kappa^{-2/3}=\gamma_0^{2/3}, \, 
	\alpha>\gamma_0^{-1/4}, 
\end{equation}
	are valid only for the acceleration of  'cold' particles, $ \gamma_i \simeq 1 $. The value of $ \gamma_0 = \omega_c / \Omega_F $ is the maximum possible value of the energy of accelerated particles (the Hillas criterion).
	When $ \gamma = \gamma_0 $, the cyclotron radius of the particle becomes equal to the radius of the light cylinder.
	In fact, the acceleration is not so strong, the powers are $ 1/2 $ and $ 2/3 $, not $ 1 $, in the real field of the magnetosphere of the rotating black hole with an accretion disk. But the presence of the jet, i.e. the toroidal magnetic field generated by an electric current flowing in a jet, makes acceleration more efficient (the power of $ 2/3 $). If particles are accelerated not from 'cold' state, but from the Lorentz factor $ \gamma_i> 1 $, then the expressions (\ref{gammam}) for them take the form
\begin{eqnarray}\label{gammami}
	&&\gamma_m\simeq(\gamma_i\gamma_0)^{1/2}, \, \alpha<(\gamma_0/\gamma_i)^{-1/4}; \nonumber \\
	&&\gamma_m\simeq\gamma_i^{1/3}\gamma_0^{2/3}, \, \alpha>(\gamma_0/\gamma_i)^{-1/4}.
\end{eqnarray}
	If particles are pre-accelerated in the black hole magnetosphere or in a turbulent accretion disk to energies $ \gamma_i >> 1 $, as it was shown by (Istomin and Sol, \hyperlink{d9}{2009)}, then the maximum energy $ \gamma_m $, obtained in the process of centrifugal acceleration, can significantly approach the value of $\gamma_0$.
	
	The dependence of $ \gamma_m $ on the angle $ \theta $, see formulas (\ref {alpha0}, \ref {alpha}), means that at different heights $ z $ above the equatorial plane the particle energy is also different.
	Accelerated particles, leaving the vicinity of the black hole, will form a single energy spectrum.
	Summing them, crossing the light surface $ r = r_L / \sin \theta $ and acquiring energy $ \gamma_m $, that depends on the angle $ \theta $, we can obtain the distribution function of fast particles $ f (\gamma_m) $. As $ z = r_L \cot \theta, \, dz = -d \theta / \sin^2 \theta $, and $ f (\gamma_m) d \gamma_m
	\propto \pi r_L^2 ndz $, where $ n $ is the density of the 'primary' particles in the magnetosphere inside the light surface ($ n \simeq const $), we get
\begin{eqnarray}\label{function}
	&&f(\gamma_m)\propto\gamma_m^{-2}, \, \gamma_m<\gamma_0^{1/2}, \,  \alpha<\gamma_0^{-1/4}; \nonumber  \\
	&&f(\gamma_m)\propto\gamma_m^{-2.5}, \, \gamma_m<\gamma_0^{2/3}, \,  \alpha>\gamma_0^{-1/4}.
\end{eqnarray}
	Here we consider that angles $ \theta $ differ from some small value of $\theta_1 $, determined by the height of the magnetosphere, to $\theta_2<\pi/2$. The angle $ \theta_2 $ separates region of the magnetosphere adjacent to the accretion disk, where the matter rotates with the speed close to the speed of rotation of the disk, and the region where the matter rotates with the speed close to the speed of rotation of the black hole. Therefore, when getting the expression (\ref {function}), one can put $ \cos \theta \simeq 1 $. The value of $ \sin \theta $ basically
	takes small values, which corresponds to the energies $ \gamma_m <\gamma_0 ^ {1/2} $ for small toroidal fields, and $ \gamma_m <\gamma_0^{2/3} $ for large ones.
			
	It should be noted that, as we have seen, for large toroidal fields, large values ​​of $ \alpha $,
	particle acceleration is more efficient than in the absence of a toroidal field. Generally speaking, there may exist such magnetic field topologies for which it is still possible
	more efficient acceleration, $ \gamma_m = \gamma_0 $. However, in case a split monopole magnetiс fields to combined with a toroidal field, the acceleration at large toroidal fields does not depend on the strength of the poloidal field, $ \gamma_m = (\alpha \kappa^{-1})^{2/3} $. It's evident  the geometry of the poloidal field that such topologies, if they exist, should rarely be realized.
    
\section{AGN}\label{section4}

Centrifugal acceleration occurs in the magnetosphere of a black hole directly in its vicinity of the size of the radius of the light cylinder, $ r \simeq r_L = c / \Omega_F $. For optimal matching of the rotation of
a magnetosphere with the rotation of a black hole the value of $ \Omega_F$ is $ \Omega_F = \Omega_H / 2 $, where $ \Omega_H $ is the angular velocity of rotation of the black hole,
\begin{eqnarray}\label{omegH}
&&\Omega_H=\frac{2c}{r_g}\frac{a}{1+(1-a^2)^{1/2}}, \nonumber  \\
&&r_g=\frac{2GM}{c^2}.
\end{eqnarray}
The value of $ r_g $ is the gravitational radius of a black hole of the $ M $ mass, $ r_g = 3 \cdot 10^5 (M / M_\odot) \, = 3 \cdot 10^{14} M_9 \, cm $. Here the value of $ M_9 $ is the mass of a black hole measured in units of $ 10^9
M_\odot $. 
We used masses of black hole$ M_9 $ given in referred papers: in the center of our Galaxy Sgr. A* (Ghez, et al., 2008) and six AGN; NGC 1365 (Fazeli, et al., 2019), NGC 4051 (Seifina, Chekhtman \& Titarchuk, 2018), NGC 4151 (Bentz, et al., 2006), NGC 4486 (M 87) (Broderick, et al., 2015), NGC 6166 (Maggorian, et al., 1998) and NGC 7469 (Seifina, Titarchuk \& Ugolkova, 2018). See the table.

Parameter $ a $ characterizes the rotation of a black hole, and is equal to its specific angular momentum, $ a <1 $. So, $\Omega_F=10^{-4}aM_9^{-1}/[1+(1-a^2)^{1/2}] \, s^{-1}$. 

The magnetic field strength in the magnetosphere $ B $ is one of characteristics of the central black hole, it plays a key role in the process of particle acceleration and underlies the calculation of $ \gamma_{m} $. The data presented in the work of Daly (2019) contain magnetic field values for some of the most famous AGN. The magnetic field strength near the horizon for six active galaxies is given by the Table \ref{tab}. The value of
$ B_4 $ is the magnitude of the magnetic field in units of $ 10^4 \, G $. 
We see that given values ​​of the magnetic field near the horizon $ B_4 $ are less than $M_9^{-1/2}$. This is due to the fact that the maximum magnetic field can be estimated as follows. Suppose that for a central object with the Eddington luminosity, $ L \simeq L_ {Edd} $, the magnetic field energy density $ B^2/8 \pi $ near the horizon is comparable with the energy density of the accreting matter (equipartition). Then
$ L_{Edd} \simeq 4 \pi cr_g^2 (B^2/8 \pi) $, and one can introduce the value of $ B_{Edd} = (2 L_{Edd} / c r_g^2)^{1 / 2} \simeq 1.7\cdot 10^4 M_9^{-1/2} \, $ G. This value, on the one hand, is the characteristic value of the magnetic field in AGN and, on the other hand, is its upper limit, $ B = bB_{Edd}, \, b <1 $. The coefficient $ b $ depends on the mass of the black hole $ M $, on the specific angular momentum of the black hole $ a $, on the kinetic luminosity of the jet $ L_j $, on the bolometric luminosity of the disk $ L_{bol}$ and on the value of the accretion rate $ {\dot M} $ . Therefore, to determine $ b $, it is necessary to know the model of the accretion disk, the jet model, etc. Using physical considerations and empirical dependences, Daly (2016, 2019) developed a procedure of determining the value of $ B_4 $, as well as $ a $.
The Daly's values ​​of the specific angular momentum $ a $ for many AGNs turned out to be close to those estimated by other methods. Thus, the magnetic fields $ B_4 $ estimated by him can be considered quite reliable.  Since the cyclotron rotation frequency for protons is equal to $ \omega_c = 10^4(B/1 \, G) \, s^{-1}$, then the value of the maximum possible Lorentz factor $ \gamma_0 =\omega_c / \Omega_F $ is 
\begin{eqnarray}\label{gam0}
	&&\gamma_0=4\cdot 10^{12}M_9B_4\frac{1+(1-a^2)^{1/2}}{a}\simeq 4\cdot 10^{12}M_9B_4, \nonumber  \\ 
	&&a\simeq 1,
\end{eqnarray}
	which corresponds the proton energy 
	$ E_ {max}^{(1)} = 3.8 \cdot 10^{21} M_9 B_4 \, eV $. However, as it follows from
	our calculations, the split monopole configuration allows to accelerating particles to values smaller than $\gamma_0$, $\gamma_m=\gamma_0^{2/3}=2.5\cdot 10^{8}(M_9B_4)^{2/3}$. For protons this is $ E_{max}^{(2/3)} = 2.3 \cdot 10^{17} (M_9B_4)^{2/3} \, eV $. In the case of a weak toroidal field, acceleration is even less efficient, $\gamma_m=\gamma_0^{1/2}=2\cdot 10^{6}(M_9B_4)^{1/2}, \, E_{max}^{(1/2)}=1.9\cdot 10^{15}(M_9B_4)^{1/2} \, eV$. 
	Here we consider the example of the galaxy M87 (NGC4486). It has a powerful jet through which electric current flows. Let us estimate the magnitudes of the poloidal and the toroidal magnetic fields in its magnetosphere. Jet luminosity is $ L = 10^{44} \, $ erg/s (Broderick et al., \hyperlink{d19}{2015)}. Equate it to the electric power released in the external part of the electric current loop. With the optimal matching, the resistance of the external electrical circuit is equal to the horizon resistance $ R_H = 4 \pi / c = 377 \, $ Ohm Thorne et al., \hyperlink{d20}{(1986)}, $ L = R_H I ^ 2 $. Then we estimate the value of the electric current $ I $ in the jet, $ I \simeq 1.6 \cdot 10 ^ {17} \, A $. This current, flowing in the jet in the forward direction in the centre and in the opposite direction at the periphery, closes in the accretion disk, flowing from the outer regions of the disc to the inner region near the black hole horizon. It creates a toroidal magnetic field $ B_\phi $ in the magnetosphere, $ B_\phi \simeq 2I / ch $, where the value of $ h $ is half the height of the disk on its inner edge. So, we get $ B_ \phi \simeq 10^2 (M_9 h / r_g)^{-1} \, G $. Given $ h / r_g \simeq 3\cdot 10^{-3} $, we find that the estimated toroidal magnetic field near the horizon is of the order of magnitude given in Table 1 for М87. Let us now estimate the value of the polodal field $ B_P $ in the vicinity of the horizon. The electric current is generated by the dynamo machine created by a rotating poloidal magnetic field. The generated voltage is $ U = B_Pr_g^2 \Omega_H / 2c $ (Landau and  Lifshitz, \hyperlink{d21}{1984}; Thorne et al., \hyperlink{d20}{1986}). On the other hand, $ U = 2R_HI $. From here we get $ B_P = 16 \pi I / cr_g = 8 \pi B_\phi (h / r_g) $. Thus, the ratio of the toroidal magnetic field to the poloidal one, i.e. $ \alpha $, equals $ \alpha = r_g / 8 \pi h \simeq 10> 1 $. Therefore, for M87 the energy of accelerated protons corresponds to the estimated value in Table \ref{tab}.

	The nucleus of our galaxy Sgr. A* contains the black hole with the mass of $ 4.3 \cdot 10^6 M_\odot $. It is not active, its luminosity is only $ \simeq 10^{-9} $ of the Eddington luminosity. There is no jet, and one can think that there is no toroidal magnetic field, $ B_\phi = 0 $. Therefore, the maximum Lorentz factor of particles is given by the expression $ \gamma_m = \gamma_0^{1/2} \simeq 8 \cdot 10^4 $, which corresponds to the proton energy of $ E_{max}^{(1/2)} \simeq 7.4 \cdot 10^{13} \, eV $. However, the observed spectrum of gamma photons from Sgr. A* is the power law with the exponent $ \simeq -2.3 $ (Abramowski et al., \hyperlink{d3}{2016)}. It is an intermediate value between the values ​​of -2 and -2.5, given by the expression (\ref{function}) for weak and strong toroidal magnetic field respectively. It should be assumed that a weak toroidal magnetic field is still present in the magnetosphere, $ \alpha \simeq \gamma_0^{-1/4} \simeq 6.3 \cdot 10^{-3} $. Plasma accretion must be accompanied by the appearance of an electric current, and hence generation of a toroidal magnetic field, the way it occurs in case of accretion onto a neutron star  (Istomin and Haensel, \hyperlink{d24}{2013)}. Therefore, the maximum possible energies of accelerated protons lie within interval $ E_{max}^{(1/2)}<E_{max} <E_{max}^{(2/3)} $, i.e. $ 7.4 \cdot 10^{13} \, eV <E_{max} <3.2 \cdot 10^{15} \, eV $, which is consistent with the observations of the Galactic centre by the HESS Cherenkov telescope (Abramowski et al., \hyperlink{d3}{2016)}.
	
	As for the power $ W $ radiated into accelerated particles, it is only part of the total power lost by the black hole, $S = R_H I^2 = U^2/4 R_H = (B_P^2/16 \pi)cr_L^2$. This part is the ratio of the maximum energy of the accelerated protons $ \gamma_m $ to the maximum possible particle energy obtained under complete transformation of the energy of rotation of the black hole into the energy of particles, $ \kappa^{-1} $, that is $ W = S \gamma_m \kappa = S \gamma_m / \gamma_0 $. For Sgr. A* the value of $ W $ lies within $ 1.5 \cdot 10^{35} $ erg/s <$ W $ <$ 7 \cdot 10^{36} $ erg/s, since $ \kappa^{-1/2 } <\gamma_m <\kappa^{-2/3} $.
		  
\begin{table}
\begin{center}
\begin{tabular}{|c|c|c|c|c|c|}
	\hline
	Galaxy & $M_9$ & $B_4$ & $E_{max}^{(1)}$ & $E_{max}^{(2/3)}$ & $E_{max}^{(1/2)}$ \\
	\hline
	Sgr. A* & 0.0043 & 0.25 & $4.1\cdot10^{18}$ & $2.4\cdot10^{15}$ & $ 6.2\cdot10^{13} $ \\
	Cyg. A* & 2.5 & 0.45 & $4.2\cdot10^{21}$ & $2.5\cdot10^{17}$ & $ 2.0\cdot10^{15} $ \\
	NGC 1365 & 0.05 & 5.0 &$9.5\cdot10^{20}$ & $9.1\cdot10^{16}$ & $9.5\cdot10^{14}$\\	
	NGC 4051 & 0.0006 & 20.0 &$4.6\cdot10^{19}$ & $1.2\cdot10^{16}$ & $2.1\cdot10^{14}$\\
	NGC 4151 & 0.05 & 4.0 &$7.6\cdot10^{20}$ & $7.9\cdot10^{16}$ & $8.5\cdot10^{14}$\\
	NGC 4258 & 0.04 & 1.45 &$2.2\cdot10^{20}$ & $3.4\cdot10^{16}$ & $4.6\cdot10^{14}$\\
	NGC 4486 & 6.6 & 0.07 & $1.7\cdot10^{21}$ & $1.4\cdot10^{17}$ & $1.3\cdot10^{15}$\\
	NGC 6166 & 28.4 & 0.11 &$1.2\cdot10^{22}$ & $5.0\cdot10^{17}$ & $3.4\cdot10^{15}$\\
	NGC 7469 & 0.003 & 20.0 &$2.3\cdot10^{20}$ & $3.5\cdot10^{16}$ & $4.7\cdot10^{14}$\\
	\hline
\end{tabular}
\end{center}
	\caption{Upper limits for proton energy. Here, $ M_9 \equiv M_{BH} / 10^9 M_\odot $, $ B_4 \equiv B_\phi / 10 ^ 4 \, G $, proton energy $E_{max}^{(1)}=m_pc^2\gamma_0,\, E_{max}^{(1/2)}= m_pc^2\gamma_0^{1/2}, \, E_{max}^{(2/3)}=m_pc^2\gamma_0^{2/3}$ given in units of $ eV $}\label{tab}
\end{table}

	Black hole masses have been updated for Sgr A∗ (Ghez et al. 2008), NGC 1365 (Fazeli et al. 2019), NGC 4051 (Seifina et al. 2018), NGC 4151 (Bentz et al. 2006), NGC 4258 Gonz{\'a}lez-L{\'o}pezlira R.~A., et al., 2017), NGC 4486 (Broderick et al. 2015), NGC 6166 (Maggorian. 1998), NGC 7469 (Seifina et al. 2018).
		
	It should be noted that we used to estimate the value of the black hole rotation parameter $a \simeq 1 $, which is true for Sgr. A* and M87. For other galaxies, the rotation may not be so fast, $ a << 1 $. In this case, the size of the magnetosphere increases, $ r_L \propto a^{-1} $, the magnetization parameter $ \kappa $ decreases by the same value, which leads to acceleration to higher energies, $ \gamma_0 \propto a^{-1}$ .
	
	We would like to emphasize that obtained estimates of $ \gamma_{m} $ are the Lorentz factor achieved by a 'cold' particle, $ \gamma_i \simeq 1 $, as a result of acceleration in the magnetosphere. If the particle has undergone preliminary acceleration, $ \gamma_i> 1 $, then the maximum possible energies increase substantially (\ref {gammam}). For example, some particles can be accelerated in a turbulent disk (Istomin and Sol, \hyperlink{d9}{2009))}.
			
	In addition to the above, we would like to add that the results obtained here can also be used for microquasars such as SS 433 (Abeysekara, \hyperlink{d23}{2018)}.
		 
\section{Discussion}\label{section5}
	
	We have shown that in the magnetosphere of the rotating black hole up to the light surface of $r <c/\Omega_F\sin\theta$, occurs an acceleration of charged particles. It is bound to the fact that the pole electric field
	$E_\theta$ makes possible the particle deviation for cyclotron radius of $ r_c = c \gamma_i/\omega_c$ from an initial magnetic surface $\theta=const$. The cyclotron radius grows in proportion to energies of a particle, $\gamma$. Therefore, acceleration is the most effective near the light surface, where $E_\theta\simeq B_r$. As a result, the work of the electric field $e c\gamma B_0/\omega_c$ is just equal to the energy derived by $mc^2\gamma$ particle. Thus, accounting of final cyclotron radius is very important for acceleration of a charged particle in a magnetosphere especially as it grows with the increase in the particle energy.
	The solution of the particle acceleration problem presented here is based on the integration of the equations of particles motion in an electromagnetic field, which is equivalent to the solution of kinetic equations for individual components - protons and electrons (we consider the electron-proton plasma in the magnetosphere rather than the electron-positron plasma, which is also possible under efficient production of pairs (Beskin, Istomin \& Pariev, \hyperlink{d26}{1992})).
	
	The problem of plasma acceleration by the rotating neutron star or black hole was also considered in the MHD approximation in many works, starting from the works of  Michel \hyperlink{d13}{(1969)}, \hyperlink{d14}{(1974)}. Here we explain in details what the MHD approximation is. In general, a neutral fluid, consisting of positive and negative charges, is described by unified equations of
	hydrodynamics type, in which the interaction of a fluid with the magnetic field is taken into account. This suggests (I) that protons and electrons move together as a single liquid, i.e. $ u_i - u_e << u_i = u + m_e (u-u_e) / m_i \simeq u $. Here, $ u $ is the average mass fluid velocity. In addition (II), the individual characteristics of the particles disappear, in particular the cyclotron radii of electrons and protons, which, generally speaking, are very different from each other, $r_{ci}>>r_{ce}$.
	
	This means that MHD equations work in regions whose characteristic size $ L $ significantly exceeds the cyclotron radius of protons, $ L >> r_{ci} $. In MHD equations, which are hydrodynamics type equations, the ion cyclotron radius plays the role of the mean free path $ l $ in ordinary hydrodynamics. Therefore, even in weakly collisional systems, $ l \simeq L $, MHD is true, if $ L >> r_{ci} $. Under ideal conditions, i.e. neglecting dissipation, the MHD approximation uses the condition of freezing of magnetic field into a liquid, $ {\bf E} = - ({\bf u}\times{\bf B}) / c $ (\ref {e}). It follows from Ohm's law $ {\bf j} = \sigma_c [{\bf E} + ({\bf u}\times{\bf B}) / c] $ under the condition of high conductivity, $ \sigma_c \to \infty $. However, the very Ohm's law, called the generalized Ohm's law (Gurnett \& Bhattacharjee, \hyperlink{d25}{2004}), obtained after protons and electrons having been considered, is as follows:

\begin{eqnarray}\label{ohm}
	{\bf j}=\sigma_c\left\{\left[{\bf E}+\frac{1}{c}[{\bf uB}]\right]+\frac{1}{enc}[{\bf jB}] +
	\frac{1}{enc}\nabla P_e\right\}.
\end{eqnarray}
	
	Here the quantity $ P_e $ is the electron pressure. Since $ {\bf j} = en ({\bf u}_i - {\bf u}_e) $, the second term on the right-hand side of the generalized Ohm's law is small compared to the first one if the condition $ {\bf u}_i - {\bf u}_e << {\bf u} \simeq {\bf u}_i $. In reality, in our problem of centrifugal particle acceleration in the black hole magnetosphere ($ r <r_L / \sin  \theta $), the polar electric field $ E_\theta $ (\ref {f}) does work on both electrons and protons, $ u_{i \theta} <0$ (Fig. \ref{ris:image3}), $ u_{e \theta}> 0 $. Therefore, $ |(u_i-u_e)_\theta | = |(u_{i \theta} - |u_{e \theta}|)|
	> |u_{i \theta}|\simeq |u_\theta| $. Thus, we see that the frozen-in condition (\ref{e}), which is one of the basic postulates of ideal MHD, is not fulfilled in our problem. It is obvious that
	if $ {\bf E} = - {\bf u}\times{\bf B} / c $, then $ {\bf E} = - {\bf u}_i\times{\bf B} / c $ with the accuracy of the small ratio $ m_e / m_i << 1 $, and the electric field does not work on protons, $ {\bf u_iE} = 0 $. In terms of physical interpreting the ideal MHD does not work here. The ideality implies not only the absence of Joule heating, but also absence of direct transfer of energy from the electric field to charged particles (the liquid is neutral). The acceleration of the flow in MHD can be only accelerated by two means: the transfer of thermal energy of the matter into kinetic energy, which is carried out in hydrodynamics, as well as the transformation of the energy of the magnetic field into the energy of the flow. This can be seen from the Euler equation in the MHD approximation:
	
\begin{eqnarray}\label{euler}
	n\frac{d{\bf p}}{dt}=-\nabla P-\frac{1}{8\pi}\nabla B^2 
	\frac{1}{4\pi}({\bf B\nabla}){\bf B}.
\end{eqnarray}  
	  
	Here, the acceleration occurs due to the pumping of the toroidal magnetic field by the central object rotation. Indeed, the radial field rapidly decreases with distance, $ \propto r^{-2} $, but the toroidal field generation, $ B_\phi \propto r^{-1} $, takes place due to the freezing-in of the magnetic field under differential rotation of the fluid. With the relation $ E_\theta = -r  \Omega_F \sin \theta B_r / c = (u_r B_\phi-u_\phi B_r) / c $, which is the $ \theta $ component of the equation (\ref{e}), arising toroidal field is
	
\begin{eqnarray}\label{bphi}
	B_\phi = B_r\frac{u_\phi-\Omega_Fr\sin\theta}{u_r}.
\end{eqnarray}
	
	Calculations in the MHD approximation show that the flow acceleration occurs at large distances from the centre, $ r >> r_L $, and up to the Lorentz factor $ \gamma_m = \sigma^{1/3} $ (Michel \hyperlink{d13}{(1969)}, \hyperlink{d14}{(1974)}; (Beskin, Kuznetsova \& Rafikov, \hyperlink{d16}{(1998)}). Here the value of $ \sigma $ is the so-called Michel's parameter, $ \sigma = e\Phi\Omega_F / mc^3 $. The flux of the radial magnetic field $\Phi$ is, $  \Phi = \pi B_0r_L^2 $. It is easy to see that $ \sigma = \pi \omega_c / \Omega_F >> 1 $. In fact, this is reciprocal to the magnetization parameter, $ \sigma \simeq \kappa^{-1}$ (\ref{kap}). It should be noted that MHD acceleration is almost isotropic, $ u_\rho \simeq u_\phi $, while the centrifugal acceleration inside the magnetosphere ($ r <r_L $) discussed in this article, is strongly anisotropic, $ p_\phi >> p_\rho, \, p_\theta $. Thus, we see that centrifugal acceleration, which is not described in the MHD approximation, is more effective for the case of a radial magnetic field (split monopole), $ \gamma_m \simeq \sigma^{1/2}, \, \sigma^{2 / 3} >> \sigma ^{1/3} $. However this is valid for the radial configuration of the magnetic field. In case are magnetic surfaces of parabolic configuration, MHD acceleration there can reach to maximum values, $ \gamma_m \simeq \sigma $ (Beskin \& Nokhrina, \hyperlink{d15}{(2006)}; Tchekovskoy, McKinney \& Narayan, \hyperlink{d17}{(2009)}). Here the acceleration is affected by the flow shape, the way it occurs in hydrodynamics (nozzle shape). However, in this case MHD calculations should be performed starting from the light surface with the boundary conditions specified by the internal flow, $ r<r_L $.	
		 
\section*{Acknowledgments}
	
	This work was supported by Russian Foundation for Fundamental Research, grant number 17-02-00788.

\end{document}